\documentclass[10pt]{iopart}
\begin{document}
\title
{Helfrich-Canham bending energy as a constrained non-linear sigma model}
\author{R  Capovilla\dag and 
J Guven\ddag}
\address{\dag\
Departamento de F\'{\i}sica,
 Centro de Investigaci\'on
y de Estudios Avanzados del IPN,
Apdo Postal 14-740, 07000 M\'exico,
D. F.,
MEXICO}
\address{\ddag
Instituto de Ciencias Nucleares,
 Universidad Nacional Aut\'onoma de M\'exico,
Apdo. Postal 70-543, 04510 M\'exico, DF, MEXICO}

\begin{abstract}
 The Helfrich-Canham  bending energy 
is identified with a non-linear sigma model for a unit vector. The 
identification, however, is dependent on one additional constraint: that
the unit vector be 
constrained to lie 
orthogonal to the surface. The presence of this constraint adds a source to 
the divergence of the stress tensor for this vector so that it is not 
conserved.  The stress tensor which is conserved is identified and 
its conservation shown to reproduce the correct shape equation.
\end{abstract}

Consider an embedded two-dimensional surface in three-dimensional
space. The positive definite geometrical invariant, 
\begin{equation}
F[{\bf X}] =
{1\over 2}
\int dA K_{ab} K^{ab}\,,
\label{eq:k2}
\end{equation}
is a measure of the energy associated with the bending of the surface.
The embedding is described by three functions ${\bf X} = X^i $ of two 
variables 
$\xi^a$, $(i,j,\dots =1,2,3; a,b,\dots =1,2)$. The extrinsic curvature 
tensor  $K_{ab}$ is then defined by 
\begin{equation}
K_{ab} =
{\bf e}_b \cdot \nabla_a {\bf n} = K_{ba} \,,
\end{equation}
where ${\bf e}_a = \partial {\bf X} / \partial \xi^a = \nabla_a {\bf X}$ are
the two 
tangent vectors to the surface 
and ${\bf n}$ is the unit normal. 
Indices are raised with
$g^{ab}$, the inverse of the induced
metric $g_{ab} = {\bf e}_a \cdot {\bf e}_b$. $dA = \sqrt{g} d^2 \xi$ is the 
area measure induced on the surface by 
${\bf X}$, with $g=$det $g_{ab}$. 

For a two-dimensional surface $F[{\bf X}]$ is  invariant 
under ambient space conformal transformations.
Since its introduction by Willmore \cite{Willmore}, 
it has cropped up in a number of contexts. 
These range from elasticity theory \cite{LL}, to cellular biophysics 
where it models the bending energy of  
phospholipid membranes and is known as the Helfrich-Canham Hamiltonian
\cite{Can,Hel,Evans},   
to its Lorentzian generalization describing the action defined on the 
two-dimensional world sheet of a relativistic string
propagating in spacetime, proposed by Polyakov to model colour flux tubes 
in QCD \cite{Polyakov1,Polyakov2,Kleinert}. We will couch our discussion
in terms of Euclidean surfaces.

The vanishing of the first variation of $F [{\bf X}]$ under an 
infinitesimal deformation of the embedding functions, ${\bf X} \to {\bf X} + 
\delta {\bf X}$, produces the shape equation \cite{Hel.OuY:87,Sei:97}
\begin{equation}
- \nabla^2 K+ {1\over 2} K(K^2 - 2 K_{ab} K^{ab})  =0\,,
\label{eq:shape}
\end{equation}
where $\nabla^2$ is the surface Laplacian, and $K = g^{ab} K_{ab}$ is the
mean extrinsic curvature. 
We remark that the shape
equation is of fourth order in derivatives of the shape functions.
Let us also note that,
a fact that will be used below, the shape equation can be written as
a conservation law \cite{Stress,auxiliary},
\begin{equation}
\nabla_a {\bf f}^a = 0\,,
\label{eq:conservation}
\end{equation}
where the stress tensor ${\bf f}^a $ is
\begin{equation}
{\bf f}^a =  K\left(K^{ab}- {1\over 2} K g^{ab} \right)\,
{\bf e}_b -  (\nabla ^a K ) \, {\bf n}\,.
\label{eq:stress}
\end{equation}

In this brief report, we would like to explore 
the fact that it is 
possible to rewrite the quadratic appearing in (\ref{eq:k2}) in the form 
\begin{equation}
K_{ab} K^{ab} = 
(\nabla_a {\bf n}) \cdot (\nabla^a {\bf n})\,,
\label{eq:kkn2}
\end{equation}
an identity which follows immediately from the definition of $K_{ab}$ 
and use of the completeness of the basis $\{{\bf e}_a, {\bf n}\}$,
$ g^{ab} e^i_a\, e^j_b = \delta^{ij} - n^i\, n^j$. This identity
associates bending energy with non-vanishing 
gradients of the normal vector;
it appears to map the bending energy
into a non-linear sigma model 
living on the curved geometry of the surface (see {\it e.g.} Ch. 
14 of \cite{ZJ}): 
\begin{equation}
F_\sigma [{\bf n}] = {1\over 2}
\int dA \; \left[ (\nabla_a {\bf n}) \cdot (\nabla^a {\bf n}) +\lambda ({\bf 
n}^2 - 1)\right] \,, \label{eq:Fsigma}
\end{equation}
with $\lambda$ a Lagrange multiplier that enforces the constraint that
${\bf n}$ be a unit vector. The 
Euler-Lagrange equations 
that follow from the vanishing of the first variation of $F_\sigma [{\bf 
n}]$ under ${\bf n} \to {\bf n} + \delta {\bf n}$ are
\begin{equation}
- \nabla^2 {\bf n} +
\lambda {\bf n} 
= 0\,,
\label{eq:elsigma}
\end{equation} 
together with the constraint ${\bf n}^2 = 1$. Using this constraint,
the Lagrange multiplier is identified as $\lambda = -
(\nabla_a {\bf n}) \cdot (\nabla^a {\bf n}) $, so that the Euler-Lagrange 
equations take the form
\begin{equation}
- \nabla^2 {\bf n} - (\nabla_a {\bf n}) \cdot (\nabla^a {\bf n}) \; {\bf n} 
= 0\,.
\end{equation} 
These equations are integrable \cite{int}; this fact is one motivation for 
investigating possible links with the bending energy (\ref{eq:k2}).

If ${\bf n}$ happens to be the unit normal to some surface, then
the fact that $\nabla_a {\bf n} = K_{ab} g^{bc} {\bf e}_c$, 
and the Codazzi-Mainardi identities $\nabla_a K^{ab} = \nabla^b K$, allow 
us to rewrite the
Euler-Lagrange equations in a geometrical form as
$ \nabla_a K = 0$.
Clearly, this is not even a distant relative of the shape equation 
(\ref{eq:shape}).
It is of second order in derivatives of ${\bf n}$, and of third
order in derivatives of ${\bf X}$. Despite  this  fact, 
unfortunately,
this erronous identification has been made frequently in the literature;
not by Polyakov, though, who was well aware of the pitfalls of a
hasty identification (see Sect. 10.4 of \cite{Polyakov2}).
What $F_\sigma [{\bf n}]$ fails to do is to capture the fact that
${\bf n}$ is normal to the surface. If ${\bf n}$  may point in
any direction, then $F_\sigma [{\bf n}]$ describes a Heisenberg 
ferromagnet on the surface. On the other hand,
if we  wish to describe the bending energy, the unit vector ${\bf n}$ must 
also satisfy the constraints
\begin{equation} {\bf n}\cdot{\bf e}_a =0
\,.
\label{eq:en}
\end{equation}
If the surface is fixed ${\bf n}$ is also; if ${\bf n}$ is 
allowed to vary, the surface must respond accordingly.
This leads to
an interesting reformulation of the bending energy.
We introduce the novel functional, with additional constraints,
\begin{eqnarray}
F[{\bf n},{\bf e}_a,{\bf X}] &=&
\int dA \, \left[ {1\over 2}
(\nabla_a {\bf n}) \cdot (\nabla^a {\bf n}) 
+
{\lambda \over 2} \, ({\bf n}^2 - 1) \right. \nonumber \\
&+& \left.  \lambda^a \, ({\bf 
n}\cdot 
{\bf e}_a)\, +\, {\bf f}^a \cdot({\bf e}_a - \nabla_a {\bf X}) \right]\,.
\label{eq:Fdef}
\end{eqnarray}
The normalization ${\bf n}^2=1$ and orthogonality constraints
(\ref{eq:en}) are implemented using the Lagrange multipliers $\lambda$ and 
$\lambda^a$, 
respectively. The latter constraints may be thought of as a frustration of 
the non-linear sigma model.
They would be simple enough to implement if the 
${\bf e}_a$ were any two fixed 
vector fields. The fact that the vectors ${\bf e}_a$ are 
the tangent vectors to the surface,
however, couples ${\bf n}$ not only to them, but also through them  
to the embedding functions ${\bf X}$ themselves; this 
connection is captured in the final set of constraints in (\ref{eq:Fdef}).
This model is a special case of a general construction introduced by one 
of us in \cite{auxiliary}, where the 
issue of implementing the necessary integrability conditions needed
for the surface to exist is sidestepped.

The Euler-Lagrange equations for ${\bf n}$, ${\bf e}_a$ and
${\bf X}$, that follow from the vanishing of the first variation
of $F[{\bf n},{\bf e}_a,{\bf X}]$,
 together 
are completely 
equivalent to the shape equation (\ref{eq:shape}). Moreover, the Lagrange
multipliers ${\bf f}^a$ appearing in (\ref{eq:Fdef}) coincide
with the stress tensor (\ref{eq:stress}) -- this justifies the abuse of
notation.
How the shape equation comes out is rather interesting.
First, the Euler-Lagrange equation for the embedding functions ${\bf X}$ 
provides 
the conservation law (\ref{eq:conservation}), $\nabla_a {\bf f}^a = 0$,
since ${\bf X}$ appears only in the last constraint.
The equations for ${\bf n}$, and ${\bf e}_a$ determine
the form of the Lagrange multipliers ${\bf f}^a$:
for ${\bf n}$ we have
\begin{equation}
 - \nabla^2 {\bf n} + \lambda {\bf n} + \lambda^a {\bf e}_a =0\,,
\label{eq:el1}
\end{equation}
and for ${\bf e}_a$, when the constraints are enforced,
\begin{equation}
{\bf f}^a = 
 T^{ab} {\bf e}_b 
- \lambda^a {\bf n} \,,
\label{eq:el2}
\end{equation}
where 
we identify the stress tensor associated with an unconstrained 
non-linear sigma model for ${\bf n}$ on 
the background intrinsic geometry of the surface
\begin{equation}
T^{ab}= (\nabla^a {\bf n}) \cdot (\nabla^b {\bf n})- {1\over 2}
g^{ab} (\nabla^c {\bf n})\cdot (\nabla_c {\bf n})\,.
\end{equation}
If $\lambda^a=0$, (\ref{eq:el1}) reproduces the Euler-Lagrange 
equations for a
non-linear sigma model (\ref{eq:elsigma}). 
The normal projection of 
(\ref{eq:el1}) gives, as before,
$
\lambda = -  (\nabla_a {\bf n}) \cdot (\nabla^a {\bf n})\,,
$
or $\lambda$ is (minus) the energy density; taking into account the 
constraint (\ref{eq:en}), its tangental counterparts give
\begin{equation}
\lambda^a = {\bf e}^a\cdot \nabla^2 {\bf n} 
= \nabla^b ({\bf e}^a\cdot \nabla_b {\bf n}) = \nabla^b K^a{}_b
= \nabla^a K\,.
\label{eq:lama}
\end{equation}
Modulo the constraints (\ref{eq:en}) and the definition of 
$K_{ab}$, we find that $T^{ab}$ can be rewritten as
\begin{equation}
T^{ab} =   K^{ac} K_{c}{}^b -{1\over 2} g^{ab} K^{cd} K_{cd}
= K\left( K^{ab} -{1\over 2} g^{ab} K \right)
\,.
\label{eq:tab}
\end{equation}
To obtain the expression on the right 
we have used the Gauss-Codazzi equation for a 
two-dimensional surface, 
\begin{equation} 
{1\over 2} {\cal R} g_{ab} = K K_{ab} - K_{ac}K^c{}_b\,,
\end{equation}
which expresses the Ricci scalar  induced by ${\bf X}$, equal to twice 
the Gaussian curvature, in terms of a quadratic in $K_{ab}$.
Therefore, inserting (\ref{eq:lama}) for $\lambda^a$ and 
(\ref{eq:tab}) for $T^{ab}$ into the Euler-Lagrange equation for ${\bf e}_a$ 
(\ref{eq:el2}), we reproduce the stress tensor (\ref{eq:stress}). 
We identify $\lambda^a$ as the normal stress and
$T^{ab}$ as the 
tangential stress in the membrane.

To arrive at the shape equation,
let us return now to the conservation law (\ref{eq:conservation}).
Using (\ref{eq:el2}), we can re-express 
the conservation law  in terms of 
tangential and normal components:
\begin{eqnarray}
 \nabla_a \lambda^a +
K^{ab} T_{ab} &=& 0\,,
\label{eq:el21}
\\
\nabla_a T^{ab} - \lambda_a K^{ab}
&=&0\,.
\label{eq:el22}
\end{eqnarray}
Using  (\ref{eq:lama}) for $\lambda^a$ and 
(\ref{eq:tab}) for $T^{ab}$, it is immediate to see that the first, normal, 
component coincides with the shape equation (\ref{eq:shape}). 
The coupling of ${\bf n}$ to the tangents provides a source to 
the sigma model stress tensor: it is no longer conserved. 
The source is the Lagrange multiplier $\lambda^a$
implementing the constraint (\ref{eq:lama}). 
When $\lambda^a$ is substituted into (\ref{eq:el22}), 
it is identically satisfied,
a feature that has its origin in the reparametrization invariance of the 
model. We note that  the Lagrange multiplier $\lambda$ does not 
appear in these equations.
We also remark on the fact that we do not need to know how
$K_{ab}$ itself responds to a deformation of the 
surface in this presentation.

What is the status of the Euler-Lagrange equation for 
${\bf n}$? Substituting $\lambda$ and $\lambda^a$ into 
(\ref{eq:el1}) 
we obtain a purely kinematical statement about 
embedded geometries identifying what 
the projections of
$\nabla^2 {\bf n}$ are with respect to the basis 
vectors, ${\bf e}_a$ and ${\bf n}$:
\begin{equation}
\nabla^2 {\bf n} = (\nabla_a K) {\bf e}^a - K^{ab}K_{ab} {\bf n} \,.
\label{eq:nabnne}
\end{equation}
To see this, 
just take a divergence of the Weingarten equations
$\nabla_a {\bf n}= K_{ab}\, {\bf e}^b$ using 
the Gauss equations $\nabla_a {\bf e}_b =-K_{ab}\,{\bf n}$ to express
$\nabla_a {\bf e}_b$ as a normal vector.

There is a  second independent quadratic invariant in the extrinsic 
curvature, involving the square of the mean curvature, 
$K^2$. However, it does not lend itself
to a simple expression of the form (\ref{eq:kkn2}). On the other hand,
the fully contracted Gauss-Codazzi equation,
\begin{equation}
{\cal R} = K^2 - K_{ab} K^{ab}\,,
\label{eq:gc1}
\end{equation}
identifies the difference between the two quadratics with 
the scalar curvature, defined intrinsically
{\it i.e.} independent of the normal ${\bf n}$.
For a two-dimensional surface,  the corresponding invariant 
is the Gauss-Bonnet topological invariant. In higher dimensions, 
the difference is still independent of ${\bf n}$. As a result, 
both $\lambda$ and $\lambda^a$ are unchanged with respect to the 
values given above. 

We note, in this context, that 
the winding number of ${\bf n}$ is 
\begin{equation}
Q= {1\over 8\pi}\int dA 
\epsilon^{ab}\epsilon_{ijk}
n^i \nabla_a  n^j \nabla_b n^k\,,
\end{equation}
where $\epsilon^{ab}$ and $\epsilon^{ijk}$ are respectively the Levi-Civita
tensors on the surface and in space \cite{kamien}.
Modulo (\ref{eq:gc1}), $Q$ 
is just the Gauss-Bonnet invariant of the surface. 
Thus once the constraint is implemented, the 
winding number is fixed by the topology of the surface.

To summarize, it has been shown 
that the coupling of a sigma model to the geometry 
constraining the unit vector ${\bf n}$ to lie normal to the surface 
converts it into the Helfrich-Canham  model --- involving only the 
surface geometry. A lot is known about both models.
One  expects  that this 
identification might allow results from one model to be 
imported into the other. In particular, one would expect this 
formulation of the Helfrich-Canham model to be 
potentially useful in statistical mechanics. 
The functional $F [{\bf X}, {\bf n}, {\bf e}_a ]$ defined by 
(\ref{eq:Fdef}) is quadratic in ${\bf n}$. This suggests that, in the
evaluation of the partition function, ${\bf n}$ may be integrated out.
This possible virtue of the formulation presented in this paper will
be the subject of future work.

\ack

We would like to thank Markus Deserno 
for helpful 
comments. JG thanks Denjoe O' Connor for useful comments and
 hospitality during his stay at DIAS.
Partial support from DGAPA-PAPIIT grant IN114302 and
from CONACyT grant 44974-F is acknowledged.

\end{document}